**THE EUROPEAN PHYSICAL JOURNAL PLUS**

Regular Article

# Locality of orthogonal product states via multiplied copies

Hao Shu[a]

College of Mathematics, South China University of Technology, Guangzhou 510641, People's Republic of China

**Abstract** In this paper, we consider the LOCC distinguishability of product states. We employ polygons to analyze orthogonal product states in any system to show that with LOCC protocols, to distinguish seven orthogonal product states, one can exclude four states via a single copy. In bipartite systems, this result implies that N orthogonal product states are LOCC distinguishable if $\lceil \frac{N}{4} \rceil$ copies are allowed, where $\lceil l \rceil$ for a real number $l$ means the smallest integer not less than $l$. In multipartite systems, this result implies that N orthogonal product states are LOCC distinguishable if $\lceil \frac{N}{4} \rceil + 1$ copies are allowed. We also give a theorem to show how many states can be excluded via a single copy if we are distinguishing n orthogonal product states by LOCC protocols in a bipartite system. Not like previous results, our result is a general result for any set of orthogonal product states in any system.

## 1 Introduction

The LOCC distinguishability of orthogonal states is one of the most important problems in quantum information theory. This property is a phenomenon of quantum locality. One of the reasons of considering such problems is that quantum channels employed in distributing states are always noisy. Therefore, users might have to confirm states they are sharing.

Let us assume that the users obtain one of the states in a set they have known. The task is to determine which state it is. The users usually have a distance such that general measurements are not allowed. However, since classical channels could be employed easily, nowadays, the different partite might consider to distinguish the states via local operations and classical commutations, that is LOCC. On the other hand, distributing multiplied number of photons in the same time might allow users to obtain multiplied copies of the state, which means that they can consume multiplied copies of the state and even destroy them in order to confirm what it is. In other words, the problem is distinguishing a set of states by LOCC via multiplied copies and can even destroy them.

There are some methods for LOCC distinguishability problems. Walgate et al. [1] gave a sufficient and necessary condition for $LPCC_1$ distinguishability of pure states and proved that any two orthogonal pure states satisfying the condition, where a LPCC protocol is a LOCC protocol in which all measurements are projective. Other methods including results of Chen et al. [2,3], Fan [4], a framework of Singal [5], Chefles [6] and Hayashi et al. [7]. There is also a result considering the relations between LOCC$_1$ indistinguishability and the dimension of the system [8].

---

[a] e-mail: Hao_B_Shu@163.com (corresponding author)

When considering LOCC distinguishability, pure states have a good property that any two orthogonal pure states are always LOCC distinguishable [1]. However, this property is not suitable for mixed states, that is, there are two orthogonal mixed states which are LOCC indistinguishable [9].

There are two special cases that authors mostly prefer to consider, of which are maximally entangled states and product states. For maximally entangled states, results such as [10–14] were given. And for product states, Bennett et al. [15] showed that an unextendible product basis is LOCC indistinguishable, on the other hand, in $C^2 \otimes C^d$, any orthogonal product states are LOCC distinguishable [15,16]. An earlier work of Chen et al. [17] stated that when distinguishing an orthogonal product basis, LOCC distinguishability is equivalent to LPCC distinguishability. Other results including constructing LOCC indistinguishable orthogonal product states such as [18–21]. In [22], there are analyses of distinguishability of orthogonal product states.

There are sets of states which are LOCC indistinguishable. This holds even for product states [23,24] which is called nolocality without entanglement. There are mainly three methods to deal with such sets. The first one is considering the unambiguous distinguishability [25–27], the second one is employing entanglement as a resource [28–31], while the third one is distinguishing states with multiplied copies [10]. Of course, distinguishing a set of LOCC indistinguishable states by other POVMs such as PPT POVMs is considerable [32,33]. However, their physical importance might less and they are out of the discussion of this article.

Let us consider the third one, distinguishing states by using LOCC protocols via multiplied copies. General (mixed) states can be indistinguishable no matter how many copies are allowed. In [9], two orthogonal mixed states which are LOCC indistinguishable with arbitrary copies are given. On the other hand, $N$ orthogonal pure states are LOCC distinguishable if $N-1$ copies are allowed, by the fact that any two orthogonal pure states are LOCC distinguishable [1]. However, the number needed for distinguishing may be less. For example, for generalized Bell states, two copies are sufficient [10].

In this paper, we consider LOCC distinguishability of orthogonal product states (which are of course pure) in bipartite or multipartite systems via multiplied copies. Hence, $n$ copies of a state means a tensor product of the state $n$ times. We prove that employing LOCC protocols (indeed only LPCC protocols), 4 of 7 states can be excluded via a single copy. This implies that $N$ orthogonal product states are LOCC distinguishable with $\lceil \frac{N}{4} \rceil$ copies in a bipartite system and with $\lceil \frac{N}{4} \rceil + 1$ copies in a multipartite system. We give a lemma to explain the difference between bipartite and multipartite systems. We also give a more generalized statement of excluding a number of states via a single copy in a bipartite system. Note that our results are independent with the dimension of the system, except the trivial restriction that the system is assumed to have $N$ orthogonal product states. We mention that distinguishability problems of product states considered by previous works are with extra settings such as for special sets [24,34] or using entanglement as a resource [35] and there are no results considering distinguishing a general set of orthogonal product states via copies without any further assumption. Therefore, our result is novel.

## 2 Results

Main results are stated in the following.

**Theorem 1** *In $C^m \otimes C^n$, to distinguish seven orthogonal product states by using LOCC protocols, a single copy is sufficient to exclude four states.*

**Theorem 2** *To distinguish N orthogonal product states in $C^m \otimes C^n$ via LOCC, $\lceil \frac{N}{4} \rceil$ copies are sufficient.*

As a corollary, we have:

**Corollary 1** *Eight orthogonal product states in $C^m \otimes C^n$ are LOCC distinguishable if two copies are allowed.*

There is a generalization of Theorem 1.

**Theorem 3** *In a bipartite system, let $S = \{|\varphi_i\rangle = |a_i\rangle|b_i\rangle | i = 1, 2, \ldots, n\}$ be a set of n orthogonal product states, and let $A = \{|a_i\rangle | i = 1, 2, \ldots, n\}$, $B = \{|b_i\rangle | i = 1, 2, \ldots, n\}$. If there exist either m states in A or m states in B which are orthogonal to each other, then by using LOCC protocols to distinguish $N \geqslant m+1$ orthogonal product states, the following statements hold:*

(1) *a single copy is sufficient to exclude m states.*
(2) *If $N \geqslant 2m + 1$, then a single copy is sufficient to exclude $m + 1$ states.*

We mention that in a bipartite system, there are nine orthogonal product states such that neither four states of Alice's partite nor four states of Bob's partite are orthogonal to each other.

Theorems 1 and 2 can be generalized to multipartite systems.

**Theorem 4** *In a multipartite system, to distinguish seven orthogonal product states by using LOCC protocols, a single copy is sufficient to exclude four states.*

**Theorem 5** *To distinguish N orthogonal product states in $\otimes_{i=1}^M C^{d_i}$ via LOCC, $\lceil \frac{N}{4} \rceil + 1$ copies are sufficient if $4|N$, and $\lceil \frac{N}{4} \rceil$ copies are sufficient, otherwise.*

The difference between Theorems 2 and 5 comes from the following lemmas.

**Lemma 1** *Four orthogonal product states in a bipartite system are always LOCC distinguishable* [22].

**Lemma 2** *There exist four orthogonal product states which are not LOCC distinguishable in a multipartite system (not trivial and with at least three partite).*

*Proof of Lemma 2* We can construct the following (unnormalized) states in a three qubits system. $|\varphi_1\rangle = |0\rangle|0\rangle|0\rangle$, $|\varphi_2\rangle = |1\rangle|0 - 1\rangle|0 + 1\rangle$, $|\varphi_3\rangle = |0 + 1\rangle|0 + 1\rangle|1\rangle$, $|\varphi_4\rangle = |0 - 1\rangle|0 + 1\rangle|1\rangle$. When distinguishing these states via LOCC protocols, by symmetry, without loss generality, assume that Alice measures firstly. Alice must measure via orthonormal basis $|0\rangle, |1\rangle$, since other partite cannot distinguish $|\varphi_1\rangle$ and $|\varphi_2\rangle$. However, if Alice's outcome is 1, then Bob and Charlie must distinguish $|\varphi_2\rangle, |\varphi_3\rangle, |\varphi_4\rangle$, which is impossible since when only considering partite of Bob and Charlie, the three states are not orthogonal. □

## 3 Proof of the results

The method using in this paper is analyzing orthogonal product states by polygons.

**Lemma 3** *Let $S = \{|\varphi_i\rangle = |a_i\rangle|b_i\rangle | i = 1, 2, \ldots, 6\}$ be a set of orthogonal product states in $C^m \otimes C^n$, $A = \{|a_i\rangle | i = 1, 2, \ldots, 6\}$, $B = \{|b_i\rangle | i = 1, 2, \ldots, 6\}$. Then there exist either three states in A or three states in B which are orthogonal to each other.*

*Proof of Lemma 3* Since $|\varphi_i\rangle$ are orthogonal to each other, we have that for $i \neq j$, either $|a_i\rangle$ is orthogonal to $|a_j\rangle$ or $|b_i\rangle$ is orthogonal to $|b_j\rangle$.

Let us mark this in a hexagon as follows. Let vertexes of the hexagon corresponding to states which are labeled by $1, 2, \ldots, 6$ and connecting vertexes i and j by a thick line if $|a_i\rangle$ is orthogonal to $|a_j\rangle$, and otherwise (and so $|b_i\rangle$ is orthogonal to $|b_j\rangle$), connecting them by a thin line.

Now any two vertexes of the hexagon are connected by a line, since product states $|\varphi_i\rangle$ are orthogonal to each other.

We will prove that there is either a thick triangle (and so there are three orthogonal states in $A$) or a thin triangle (and so there are three orthogonal states in $B$) in the hexagon.

The hexagon has totally 15 lines connecting any two vertexes. Thus, there are at least eight thick lines or at least eight thin lines. Without loss generality, assume that there are at least eight thick lines. If every vertex is connected by at most two thick lines, then it will be at most six thick lines, and here is not such case. Therefore, there is a vertex which is connected by at least three thick lines. Without loss generality, assume that vertex 1 has thick lines connecting with vertexes 2, 3, 4. Please see Graph 1 (a).

If there is a thick line connecting either vertexes 2 and 3, vertexes 2 and 4, or vertexes 3 and 4, then there is a thick triangle. Otherwise, there are thin lines connecting vertexes 2, 3 and 4, and then there is a thin triangle. Please see Graph 1 (b). □

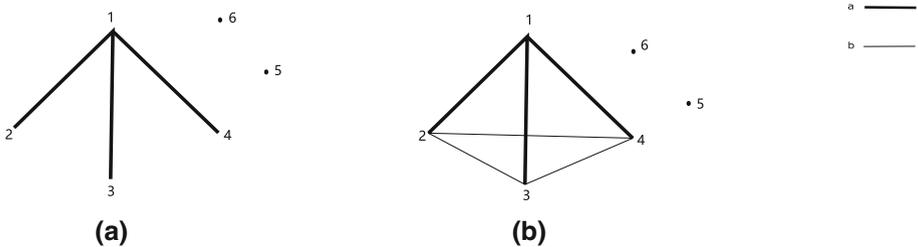

Graph 1

*Proof of Theorem 1* Let seven orthogonal product states in $C^m \otimes C^n$ be $|\varphi_i\rangle = |a_i\rangle|b_i\rangle$, $i = 1, 2, \ldots, 7$. Using Lemma 3, without loss generality, assume that $|a_1\rangle, |a_2\rangle, |a_3\rangle$ are orthogonal to each other. There are three cases.

**Case 1**: There are five $|a_i\rangle$ orthogonal to each other. Without loss generality, assume that $|a_i\rangle$, $i = 1, 2, \ldots, 5$ are orthogonal to each other.

Now, Alice measures her partite via a normalized orthogonal basis completed by $|a_i\rangle$, $i = 1, 2, \ldots, 5$ and then she can exclude at least four states (if the outcome is not 1, 2, 3, 4, 5, then $|\varphi_i\rangle$, $i = 1, 2, \ldots, 5$ are excluded, and if the outcome is one of 1, 2, 3, 4, 5, then other four states are excluded).

**Case 2**: There are four $|a_i\rangle$ orthogonal to each other. Without loss generality, assume that $|a_i\rangle$, $i = 1, 2, \ldots, 4$ are orthogonal to each other.

Now, Alice measures her partite via a normalized orthogonal basis completed by $|a_i\rangle$, $i = 1, 2, \ldots, 4$ and gets an outcome, says j.

Case 2.1: $j \neq 1, 2, 3, 4$. In this case, $|\varphi_i\rangle$, $i = 1, 2, \ldots, 4$ are excluded.

Case 2.2: $j \in \{1, 2, 3, 4\}$. Without loss generality, assume that $j = 4$, and so Alice excludes $|\varphi_i\rangle$, $i = 1, 2, 3$. If $|a_4\rangle$ is orthogonal to $|a_5\rangle$, then $|\varphi_5\rangle$ is also excluded. If it is not such case, says $|a_4\rangle$ is not orthogonal to $|a_5\rangle$, then $|b_4\rangle$ is orthogonal to $|b_5\rangle$. Now Bob measures his partite via a normalized orthogonal basis completed by $|b_4\rangle$ and $|b_5\rangle$, then he can exclude either $|\varphi_4\rangle$ or $|\varphi_5\rangle$. In both cases, they totally exclude at least four states.

**Case 3**: There are three $|a_i\rangle$ which are orthogonal to each other but no four $|a_i\rangle$ which are orthogonal to each other. Without loss generality, assume that $|a_i\rangle$, $i = 1, 2, 3$ are orthogonal to each other.

Now, Alice measures her partite via a normalized orthogonal basis completed by $|a_i\rangle$, $i = 1, 2, 3$ and gets an outcome, says j.

Case 3.1: $j \neq 1, 2, 3$. In this case, $|\varphi_i\rangle$, $i = 1, 2, 3$ are excluded. Now, no four $|a_i\rangle$ are orthogonal to each other which implies that there exist $l \neq k$, where $l, k \geqslant 4$ such that $|a_l\rangle$ is not orthogonal to $|a_k\rangle$, and so $|b_l\rangle$ is orthogonal to $|b_k\rangle$. Now Bob measures his partite via a normalized orthogonal basis completed by $|b_l\rangle$ and $|b_k\rangle$, then he can exclude either $|\varphi_l\rangle$ or $|\varphi_k\rangle$. Now, they totally exclude at least four states.

Case 3.2: $j \in \{1, 2, 3\}$. Without loss generality, assume that $j = 3$, and so Alice excludes $|\varphi_i\rangle$, $i = 1, 2$.

Case 3.2.1: $|a_3\rangle$ is orthogonal to two $|a_k\rangle$, $k \geqslant 4$. Then such two $|\varphi_k\rangle$ are also excluded. And so Alice excludes four states.

Case 3.2.2: $|a_3\rangle$ is orthogonal to a unique $|a_k\rangle$, $k \geqslant 4$. Then such $|\varphi_k\rangle$ can also be excluded. Without loss generality, assume that $|\varphi_4\rangle$ is excluded. Now $|a_l\rangle$ is not orthogonal to $|a_3\rangle$, $l = 5, 6, 7$. And so $|b_l\rangle$ is orthogonal to $|b_3\rangle$, $l = 5, 6, 7$. Let Bob measures his partite via a normalized orthogonal basis completed by $|b_3\rangle$ and $|b_5\rangle$, then he can exclude either $|\varphi_3\rangle$ or $|\varphi_5\rangle$. Now they totally exclude at least four states.

Case 3.2.3: No $|a_k\rangle$ are orthogonal to $|a_3\rangle$, $k \geqslant 4$. Then $|b_3\rangle$ is orthogonal to $|b_k\rangle$, $k = 4, 5, 6, 7$. Now, no four $|a_i\rangle$ are orthogonal to each other which implies that there exist $l \neq k$, where $l, k \geqslant 4$ such that $|a_l\rangle$ is not orthogonal to $|a_k\rangle$, and so $|b_l\rangle$ is orthogonal to $|b_k\rangle$. Now B measures his partite via a normalized orthogonal basis completed by $|b_3\rangle$, $|b_l\rangle$ and $|b_k\rangle$, then he can exclude two of $|\varphi_3\rangle$, $|\varphi_l\rangle$ and $|\varphi_k\rangle$. Thus, they can totally exclude at least four states.

Above discussions have contained all possible cases and so they can exclude at least four states via LOCC protocols with a single copy of the state. □

Now, by using Theorem 1 and Lemma 1, let us prove Theorem 2.

*Proof of Theorem 2* Write $N = 4k + r$, where $k, r$ are nonnegative integers and $r = 0, 1, 2, 3$. If $k \geqslant 2$, then we can exclude four states via a single copy, by using Theorem 1. And so we can use $k - 1$ copies to exclude $4(k - 1)$ states and left $4 + r$ states.

If $r = 0$, then by using Lemma 1, the left states are distinguishable via LOCC by a single copy.

If $r \geqslant 1$, by using Lemma 1, at least three states can be excluded via a single copy, leaving at most four states. Then using Lemma 1 again, the left states can be distinguished via LOCC by another single copy.

In both cases, the total number of copies needed are at most $\lceil \frac{N}{4} \rceil$. □

The proof of Theorem 3 is similar to the proof of Theorem 1, while the proof of Theorem 5 is similar to the proof of Theorem 2, by using the fact that three orthogonal product states in a multipartite system are always LOCC distinguishable.

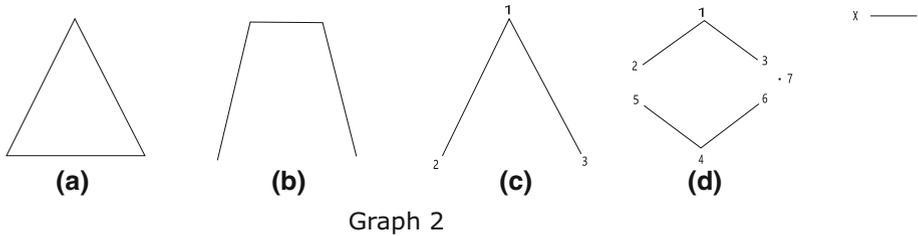

Graph 2

*Proof of Theorem 5* Write $N = 4k + r$, where $k, r$ are non-negative integers and $r = 0, 1, 2, 3$. If $k \geq 2$, then we can exclude four states via a single copy, by Theorem 4. Thus, by $k - 1$ copies, $4(k - 1)$ states can be excluded and left $4 + r$ states.

Now, at most two more copies are needed for the reason that employing another single copy can leave at most three states which are distinguishable via another. In more details, there are at most seven states left and if the number of states is less than seven, we can add some states to seven. Now, four states can be excluded via the first copy and others are distinguishable via the other.

The total number of copies needed to be used are at most $\lceil \frac{N}{4} \rceil + 1$ when $4|N$ and $\lceil \frac{N}{4} \rceil$, otherwise. □

We still have to prove Theorem 4. The proof is based on analyzing heptagons. Let us prove it for a tripartite system. The proof for other cases will be given in the "Appendix".

Firstly, we prove two lemmas.

**Lemma 4** *Using three colors, a, b, c, to color all twenty-one edges (including diagonals) of a heptagon, there exists a color $x \in \{a, b, c\}$ such that there is a subgraph of form (a) or (b) in Graph 2 with all edges are colored by x.*

*Proof of Lemma 4* There are 21 edges totally and so at least one kind of colors, says x, color seven edges. If there are no subgraphs of form (a) and (b) in Graph 2, then there are three cases.

Case 1: Every vertex is connected by at most one x-colored edge and so there are at most three x-colored edges.

Case 2: There is a unique vertex, says 1, be a vertex connected by more than one x-colored edges. Without loss generality, there is a subgraph of form (c) in Graph 2. Now there are no x-colored edges with a vertex be 2 or 3, and there are at most four x-colored edges with a vertex be 4, 5, 6, 7. Thus, there are at most six x-colored edges.

Case 3: There are two vertexes connected by two x-colored edges. Without loss generality, there is a subgraph of form (d) in Graph 2. Now there are no other x-colored edges with a vertex be 2, 3, 5, 6, since it is no of form (a) (b) of Graph 2. For there are seven x-colored edges, vertexes 1, 4, 7 must be all connected by x-colored edges, contradict with that there are no subgraphs of form (a) in Graph 2. □

**Lemma 5** *In a tripartite system, for a set of seven orthogonal product states $S = \{|\varphi_i\rangle = |a_i\rangle|b_i\rangle|c_i\rangle \ldots |i = 1, 2, \ldots, 7\}$. Let $A = \{|a_i\rangle|i = 1, 2, \ldots, 7\}$, $B = \{|b_i\rangle|i = 1, 2, \ldots, 7\}$, $C = \{|c_i\rangle|i = 1, 2, \ldots, 7\}$, there exists a partite X=A,B,C such that either of the following statements hold:*

(1) *There are three states in X which are orthogonal to each other.*
(2) *There are four states in X, say, $|x_{i_1}\rangle, |x_{i_2}\rangle, |x_{i_3}\rangle, |x_{i_4}\rangle$ such that the three pairs $|x_{i_1}\rangle$ and $|x_{i_2}\rangle$, $|x_{i_1}\rangle$ and $|x_{i_3}\rangle$, $|x_{i_2}\rangle$ and $|x_{i_4}\rangle$ are orthogonal.*

*Proof of Lemma 5* Let us mark orthogonal relations in a heptagon with vertexes corresponding to states while there is a x-colored edge connect vertexes i and j if $|x_i\rangle$ and $|x_j\rangle$ are orthogonal. Then Lemma 5 is the same as Lemma 4. □

For convenience, let us use the following notions in the rest of the paper.

**Notions**:

(1) We use H to denote the system and H is assumed to be finite dimensional. $S = \{|\varphi_i\rangle = |a_i\rangle|b_i\rangle|c_i\rangle \ldots |i = 1, 2, \ldots, 7\}$ is assumed to be a set of seven orthogonal product states in H. $X = \{|x_i\rangle | i = 1, 2, \ldots, 7\}$, that is a capital letter $X = A, B, C, \ldots$ denotes a partite and also the set of states of the partite, while the corresponding small letter with subscripts $|x_i\rangle = |a_i\rangle, |b_i\rangle, |c_i\rangle, \ldots$ denotes a state in X.
(2) If $\{|x_{i_1}\rangle, |x_{i_2}\rangle, \ldots\}$ is an orthonormal set, then we say "X measures by $_xM_{i_1,i_2,\ldots}$", "measures by $_xM_{i_1,i_2,\ldots}$", or even simply "$_xM_{i_1,i_2,\ldots}$" by means of "Partite X provides a local measurement via an orthonormal basis extended by $|x_{i_1}\rangle, |x_{i_2}\rangle \ldots$". For example, if $|a_1\rangle \perp |a_2\rangle$, then "$_aM_{1,2}$" is of the meaning that "A provides a local measurement via an orthonormal basis extended by $|a_1\rangle, |a_2\rangle$," where we also use the orthogonal symbol "$\perp$". On the other hand, after $_xM_{i_1,i_2},\ldots$, we say the outcome is 0, if the outcome is not one of $i_1, i_2, \ldots$.
(3) We use "WLG" for "Without loss generality," and we say two orthogonal product states are orthogonal via X if the X partite of the two states are orthogonal.
(4) We may use a heptagon to mark orthogonal relations of S with vertexes corresponding to states and x-edges corresponding to orthogonal relations via partite X.
(5) We use notion "i__x__j" by means of vertexes i and j are connected by an edge with x-colored or equivalently states $|\varphi_i\rangle$ and $|\varphi_j\rangle$ are orthogonal via partite X.

Now, let us prove Theorem 4 for a tripartite system.

*Proof of Theorem 4 for a tripartite system*: By symmetry, WLG, we have the following cases.

**Case 1**: There are four states in a partite, which are orthogonal to each other. WLG, assume that $\{|a_i\rangle | i = 1, 2, 3, 4\}$ is an orthonormal set. Then $_aM_{1,2,3,4}$ and gets an outcome j.

Case 1.1: $j = 0$, then the four states are excluded.

Case 1.2: WLG, $j = 4$ and so $|\varphi_1\rangle, |\varphi_2\rangle, |\varphi_3\rangle$ are excluded. If $|a_4\rangle \perp |a_5\rangle$, then $|\varphi_5\rangle$ is excluded; If not, then WLG, $|b_4\rangle \perp |b_5\rangle$, and so $_bM_{4,5}$ can exclude another state.

**Case 2**: There are three states in a partite, which are orthogonal to each other while no four states in a partite which are orthogonal to each other. WLG, assume that $\{|a_i\rangle | i = 1, 2, 3\}$ is an orthonormal set. Then $_aM_{1,2,3}$ and gets an outcome j.

Case 2.1: $j = 0$, and so three states are excluded. Now $\{|a_i\rangle | i = 4, 5, 6, 7\}$ is not an orthogonal set. Thus, WLG, $|b_4\rangle \perp |b_5\rangle$ and so $_bM_{4,5}$ can exclude another state.

Case 2.2: WLG, $j = 3$ and so $|\varphi_1\rangle, |\varphi_2\rangle$ are excluded.

Case 2.2.1: In $\{|a_i\rangle | i = 4, 5, 6, 7\}$, if there are two states which are orthogonal to $|a_3\rangle$, then they are excluded.

Case 2.2.2: In $\{|a_i\rangle | i = 4, 5, 6, 7\}$, if there is exactly one state, WLG, $|a_4\rangle$, which is orthogonal to $|a_3\rangle$, then $|a_3\rangle$ is not orthogonal to $|a_5\rangle$. WLG, assume that $|b_3\rangle \perp |b_5\rangle$. Now, $_bM_{3,5}$ and then they can totally exclude at least 4 states.

Case 2.2.3: In $\{|a_i\rangle | i = 4, 5, 6, 7\}$, no states are orthogonal to $|a_3\rangle$. Since $\{|a_i\rangle | i = 4, 5, 6, 7\}$ is not an orthogonal set, WLG, $|b_4\rangle \perp |b_5\rangle$

Case 2.2.3.1: $|b_4\rangle \perp |b_3\rangle$, $|b_3\rangle \perp |b_5\rangle$, then $_bM_{3,4,5}$.

Case 2.2.3.2: $|b_3\rangle$ is orthogonal to one of $|b_4\rangle$, $|b_5\rangle$ and WLG, $|b_4\rangle \perp |b_3\rangle$ so that $|c_3\rangle \perp |c_5\rangle$. Then $_bM_{4,5}$ and gets an outcome t. If $t = 0$, then $|\varphi_4\rangle$, $|\varphi_5\rangle$ are excluded. If $t = 5$, then $_cM_{3,5}$ can exclude another state. If $t = 4$, then $|\varphi_3\rangle$ and $|\varphi_5\rangle$ are excluded.

Case 2.2.3.3: $|b_3\rangle$ is not orthogonal to $|b_4\rangle$ and $|b_5\rangle$, then $|c_4\rangle \perp |c_3\rangle$, $|c_3\rangle \perp |c_5\rangle$. Now, $_bM_{4,5}$ and gets an outcome t. If $t = 0$, then $|\varphi_4\rangle$, $|\varphi_5\rangle$ are excluded. If $t = 5$, then $_cM_{3,5}$. If $t = 4$, then $_cM_{3,4}$.

In all cases, at least four states can be excluded totally.

**Case 3**: If no three states in a partite form an orthogonal set. By Lemma 4, WLG, assume that $|a_1\rangle \perp |a_2\rangle$, $|a_1\rangle \perp |a_3\rangle$, $|a_2\rangle \perp |a_4\rangle$ and $|a_1\rangle$ is not orthogonal to $|a_4\rangle$ while $|a_2\rangle$ is not orthogonal to $|a_3\rangle$. Now, $_aM_{1,2}$ and gets an outcome $j$.

Case 3.1: $j = 0$. Thus, $|\varphi_1\rangle$, $|\varphi_2\rangle$ are excluded.

Case 3.1.1: In $\{|a_i\rangle | i = 3, 4, 5, 6, 7\}$, there exists a state, WLG, $|a_4\rangle$, which is orthogonal to three states, WLG, $|a_5\rangle$, $|a_6\rangle$, $|a_7\rangle$. Now, every pair of $|a_5\rangle$, $|a_6\rangle$, $|a_7\rangle$, is not orthogonal, and so WLG, $|b_5\rangle \perp |b_6\rangle$, $|b_5\rangle \perp |b_7\rangle$, $|c_6\rangle \perp |c_7\rangle$. Then $_bM_{5,6}$ and gets an outcome t. If $t = 0$, then $|\varphi_5\rangle$, $|\varphi_6\rangle$ are excluded. If $t = 5$, then $|\varphi_6\rangle$ and $|\varphi_7\rangle$ are excluded. If $t = 6$, then $_cM_{6,7}$.

Case 3.1.2: In $\{|a_i\rangle | i = 3, 4, 5, 6, 7\}$, no state is orthogonal to three states. Since $\{|a_i\rangle | i = 3, 4, 5\}$ is not an orthogonal set, WLG, $|b_3\rangle \perp |b_4\rangle$. Now, $_bM_{3,4}$ and gets an outcome t. If $t = 0$, then $|\varphi_3\rangle$, $|\varphi_4\rangle$ are excluded. If $t = 3$ or $4$, WLG, $t = 3$ and then $|\varphi_4\rangle$ is excluded. If moreover, $|b_3\rangle \perp |b_i\rangle$, for some $i = 5, 6, 7$, then $|\varphi_i\rangle$ is excluded. If not, then since $|a_3\rangle$ is not orthogonal to all $|a_l\rangle$, for $l = 5, 6, 7$, WLG, $|a_3\rangle$ is not orthogonal to $|a_5\rangle$ and since $|b_3\rangle$ is not orthogonal to $|b_5\rangle$, we have $|c_3\rangle \perp |c_5\rangle$. Now, $_cM_{3,5}$.

In all above cases, at least four states can be excluded.

Case 3.2: WLG, $j = 2$. Thus, $|\varphi_1\rangle$, $|\varphi_4\rangle$ are excluded.

Case 3.2.1: In $\{|a_i\rangle | i = 5, 6, 7\}$, there are two states, WLG, $|a_5\rangle$ and $|a_6\rangle$, which are orthogonal to $|a_2\rangle$, then $|\varphi_5\rangle$ and $|\varphi_6\rangle$ are excluded.

Case 3.2.2: In $\{|a_i\rangle | i = 5, 6, 7\}$, there is a unique state, WLG, $|a_5\rangle$ which is orthogonal to $|a_2\rangle$, then WLG, $|b_2\rangle$ is orthogonal to $|b_6\rangle$. Then $_bM_{2,6}$.

Case 3.2.3: In $\{|a_i\rangle | i = 5, 6, 7\}$, no states are orthogonal to $|a_2\rangle$. Since the above set is not an orthogonal set, WLG, assume that $|a_5\rangle$ is not orthogonal to $|a_6\rangle$ and so $|b_5\rangle \perp |b_6\rangle$.

Case 3.2.3.1: $|b_2\rangle \perp |b_5\rangle$, $|b_2\rangle \perp |b_6\rangle$, then it is of case 2 or case 1.

Case 3.2.3.2: $|b_2\rangle$ is orthogonal to exactly one state of $|b_5\rangle$ and $|b_6\rangle$ and WLG, $|b_2\rangle \perp |b_5\rangle$ and then $|c_2\rangle \perp |c_6\rangle$. Now $_bM_{5,6}$ and gets an outcome t. If $t = 0$, then $|\varphi_5\rangle$ and $|\varphi_6\rangle$ are excluded. If $t = 5$, then $|\varphi_2\rangle$ and $|\varphi_6\rangle$ are excluded. If $t = 6$, then $_cM_{2,6}$.

Case 3.2.3.3: $|b_2\rangle$ is not orthogonal to any state of $|b_5\rangle$ and $|b_6\rangle$, then $|c_2\rangle \perp |c_5\rangle$, $|c_2\rangle \perp |c_6\rangle$. Now, $_bM_{5,6}$ and gets an outcome t. If $t = 0$, then $|\varphi_5\rangle$ and $|\varphi_6\rangle$ are excluded. If $t = 5$ or $6$, then $_cM_{2,t}$.

In all above cases, at least four states can be excluded.

We have discussed all cases. □

## 4 Example

In this section, let us give an example for distinguishing eight orthogonal product states via LOCC by using two copies in a bipartite system.

The nine domino states (unnormalized) in [2] form an orthogonal product basis of $C^3 \otimes C^3$. They are $|\varphi_{1,2}\rangle = |0\rangle|0 \pm 1\rangle$, $|\varphi_{3,4}\rangle = |0 \pm 1\rangle|2\rangle$, $|\varphi_{5,6}\rangle = |2\rangle|1 \pm 2\rangle$, $|\varphi_{7,8}\rangle = |1 \pm 2\rangle|0\rangle$, $|\varphi_9\rangle = |1\rangle|1\rangle$. These states are LOCC indistinguishable even if omitting $|\varphi_9\rangle$. We will use

the protocol of Theorem 1 to show that $|\varphi_i\rangle$, $i = 1, 2, \ldots, 8$, are LOCC distinguishable via two copies.

Alice measures via the computational basis on the first copy. If the outcome j is 0, then $|\varphi_i\rangle$, $i = 5, 6, 7, 8$ are excluded. If j is 1, then $|\varphi_i\rangle$, $i = 1, 2, 5, 6$ are excluded. If j is 2, then $|\varphi_i\rangle$, $i = 1, 2, 3, 4$ are excluded. Then by using the other copy, they can distinguish other four states via LOCC. In fact, since Alice excluded four states before Bob's measurement, and after Alice's measurement, there exist two unexcluded orthogonal states of Bob's partite, Bob can exclude some states by a local measurement. In the example, using a single copy, they can exclude six states. We note that in case 1 and case 2 of Theorem 1, one can exclude more than four states.

## 5 Conclusion and discussion

In this paper, we discuss the locality of orthogonal product states via multiplied copies. We prove that to distinguish seven orthogonal product states via LOCC protocols, one can exclude four states via a single copy. We also give a more generalized statement for bipartite system. Using the theorem, we give a theorem that one can distinguish N orthogonal product states via LOCC with $\lceil \frac{N}{4} \rceil + 1$ copies.

The main method we use in this paper is marking the orthogonal relations of a set of orthogonal product states in a polygon and then analyzing polygons. It is interesting and significant and problems become very mathematical.

Instead of considering distinguishability problems for certain sets of product states, we consider the problem for a general set of product states, while instead of using extra resources such as entanglement, we consider distinguishability problems of product states via multiplied copies. Our results are better than previous results. For examples, the result in [1] can imply that one can use $N-1$ copies to distinguish N orthogonal pure states in a multipartite system, and [22] can imply that one can use three copies to distinguish eight orthogonal product states in a bipartite system. Our results, on one hand, consume less states and, on the other hand, are suitable even in multipartite systems. Moreover, they are independent of the dimension of system except the nature restriction. And so far, for the results we have known, there are no authors considered such a problem for general sets of product states.

The problems left are mainly mathematical. For example, using thick lines and thin lines to connect all vertexes of a n-polygon, and finding the maximal number m such that there exist m vertexes with lines between them are all thick or thin. This will give conditions of Theorem 3. Another interesting direction for further discussions is considering quantum theory together with graph theory. As we have seen, several relations of states can be marked in a graph.

**Author contributions** The article has a unique author.

**Funding** No funding.

**Data availability** The author declares that all data supporting the findings of this study are available within the paper.

**Declarations**

**Competing interests** The author declares no competing interests.

**Code availability** No needed.

## Appendix

*Proof of Theorem 4 for a system with at least four partite*: The proof is similar to tripartite case but have other cases of following.

(1) In case 2.2.3.3, there is another case. Instead of $|c_3\rangle \perp |c_4\rangle$, and $|c_3\rangle \perp |c_5\rangle$, WLG, assume that $|c_3\rangle \perp |c_4\rangle$, and $|d_3\rangle \perp |d_5\rangle$.

In such case, the discussion becomes $_bM_{4,5}$ and gets an outcome t. If t=0, then $|\varphi_4\rangle, |\varphi_5\rangle$ are excluded. If t=4, then $_cM_{3,4}$. If t=5, then $_dM_{3,5}$.

(2) In case 3.1.1, there is another case. Instead of $|b_5\rangle \perp |b_6\rangle$, $|b_5\rangle \perp |b_7\rangle$, $|c_6\rangle \perp |c_7\rangle$, WLG, assume that $|b_5\rangle \perp |b_6\rangle$, $|c_5\rangle \perp |c_7\rangle$, $|d_6\rangle \perp |d_7\rangle$.

In such case, the discussion becomes $_bM_{5,6}$ and gets an outcome t. If $t = 0$, then $|\varphi_5\rangle, |\varphi_6\rangle$ are excluded. If $t = 5$, then $_cM_{5,7}$. If $t = 6$, then $_dM_{6,7}$.

(3) **Case 4**: There is no partite satisfying Case 1, Case 2, and Case 3.

**Case 4.1**: WLG, there is a subgraph of form (a) in Graph 3. Then there are two subcases.
Case 4.1.1: WLG, there is a subgraph of form (b) in Graph 3.
Case 4.1.2: WLG, there is a subgraph of form (c) in Graph 3.

We use tables to give protocols of such cases, note that "$_xM_{i_1,i_2,\ldots}$" now means that "WLG, we can assume that $|x_{i_1}\rangle, |x_{i_2}\rangle, \ldots$ are orthonormal and the partite X provides a local measurement via an orthonormal basis extended by $|x_{i_1}\rangle, |x_{i_2}\rangle, \ldots$." This notion is also used in the rest of the "Appendix".

**Case 4.2**: There are no subgraphs of form of Case 4.1, but has a subgraph of one of forms (a), (b), (c),(d), in Graph 4.

**Case 4.3**: Other cases (Graph 5)

The above are all cases and the proof is finished. □

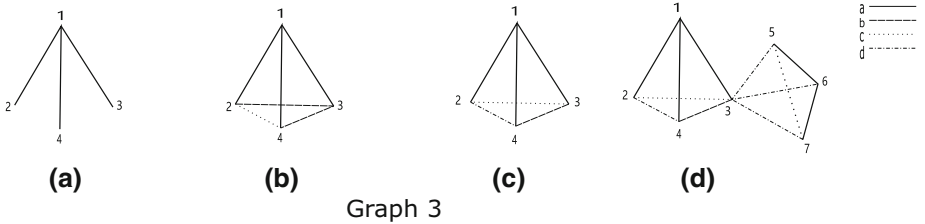

Graph 3

| Table 4.1.1 (Graph 3 (b)) | | | | | | | |
|---|---|---|---|---|---|---|---|
| | outcome | operate | outcome | operate | outcome | operate |
| $_aM_{1,2}$ | 0 | $_bM_{3,4}$ | 0 | | | | |
| | | | 3 (If the outcome is 3 or 4, WLG, assume that it is 3.) | If 3 $_b$ 5, then 5 is excluded; If not, since it is not case 3, then $_cM_{3,5}$. | | |
| | 1 | If 1 $_a$ 5, then 5 is excluded; If not, then $_bM_{1,5}$. | | | | | |
| | 2 | $_bM_{2,3}$ | 0 | If it is not case 3, then $_cM_{4,5}$. | | |
| | | | 2 | $_cM_{2,4}$ | 0 | |
| | | | | | 2 (If the outcome is 2 or 4, WLG, assume that it is 2.) | If 2 $_c$ 5, then 5 is excluded; If not, since it is not case 3, then $_dM_{2,5}$. |
| | | | 3 | If 3 $_b$ 5, then 5 is excluded; If not, since it is not case 3, then $_dM_{3,5}$. | | |

Table 4.1.2(Graph 3 (c))

| outcome | operate | outcome | operate | outcome | operate |
|---|---|---|---|---|---|
| $_aM_{1,2}$ | 0 | $_bM_{3,4}$ | 0 | $_cM_{3,5}$ | | |
| | | | 3 (If the outcome is 3 or 4, WLG, assume that it is 3.) | | | |
| | 1 | If 1 $\underline{a}$ 5, then 5 is excluded; If not, then $_aM_{1,}$. | | | | |
| | 2 | $_dM_{2,4}$ | 0 | If 3 $\underline{x}$ i, for some i=5,6,7 and x≠d, since it is not case 3, then $_bM_{3,5}$; If not, then since it is not case 2 or 3, WLG, we can assume that 5 $\underline{x}$ 7 such that x≠a,d. Now $_cM_{5,7}$. | | |
| | | | | | 0 | |
| | | | | | 2 | If 2 $\underline{x}$ 5, with x=a,c,d, then 5 is excluded; If not, then $_dM_{2,5}$. |
| | | | 2 (If the outcome is 2 or 4, WLG, assume that it is 2.) | $_cM_{2,3}$ | 3 | If i $\underline{x}$ j, for some i,j=5,6,7 and x≠a,b,c, then $_dM_{5,6}$; If 3 $\underline{c}$ i for some i≥5, then i is excluded; If 3 $\underline{x}$ i for some i≥5, and x ≠c,d, since it is not case 3, then $_bM_{3,5}$; If not such cases, since it is not case 3, then 3 $\underline{d}$ i for all i≥5, and i $\underline{x}$ j, for i,j=5,6,7 imply that x=a,c. (Please see Graph 3 (d)) Hence, it is of case 4.1.1. |

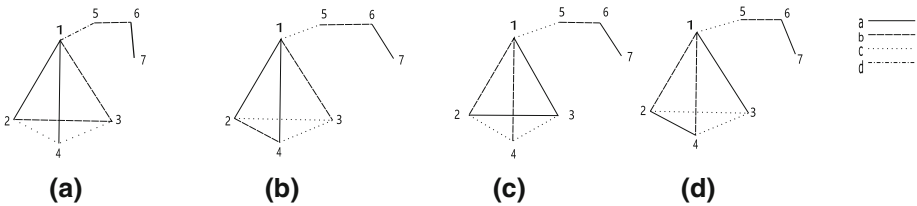

Graph 4

| | Table 4.2 (Graph 4 (a)) | | | |
|---|---|---|---|---|
| | outcome | operate | outcome | operate |
| $_aM_{1,2}$ | 0 | $_cM_{3,4}$ | 0 | |
| | | | 3 | $_bM_{5,6}$ |
| | | | 4 | |
| | 1 | $_cM_{1,5}$ | 0 | |
| | | | 1 | $_bM_{1,3}$ |
| | | | 5 | $_bM_{5,6}$ |
| | 2 | $_cM_{2,4}$ | 0 | $_bM_{5,6}$ |
| | | | 2 | For 2 $\underline{x}$ 7, x can not be a,b,c, or it becomes case 3. And so we can measure $_dM_{2,7}$, and then $_bM_{5,6}$. |
| | | | 4 | $_bM_{5,6}$ |

| Table 4.2(Graph 4 (b)) | | | | | | | |
|---|---|---|---|---|---|---|---|
| outcome | operate | outcome | operate | outcome | operate | | |
| $_aM_{1,2}$ | 0 | $_cM_{3,4}$ | 0 | | | |
| | | | 3 | $_bM_{5,6}$ | | |
| | | | 4 | | | |
| | 1 | $_cM_{1,5}$ | 0 | | | |
| | | | 1 | $_bM_{1,3}$ | | |
| | | | 5 | $_bM_{5,6}$ | | |
| | 2 | $_cM_{2,3}$ | 0 | $_bM_{5,6}$ | | |
| | | | 2 | For 2_x_6, x can not be a,b,c, or it becomes case 3. And so $_dM_{2,6}$. | 0 | |
| | | | | | 2 | $_bM_{2,4}$ |
| | | | | | 6 | $_bM_{5,6}$ |
| | | | 3 | $_bM_{5,6}$ | | |

| Table 4.2(Graph 4 (c)) | | | | | | | |
|---|---|---|---|---|---|---|---|
| outcome | operate | outcome | operate | outcome | operate | | |
| $_aM_{1,3}$ | 0 | $_cM_{2,4}$ | 0 | | | |
| | | | 2 | $_bM_{5,6}$ | | |
| | | | 4 | | | |
| | 1 | $_cM_{2,4}$ | 0 | $_bM_{5,6}$ | | |
| | | | 2 | For 2_x_6, x can not be a,b,c, or it becomes case 3. And so $_dM_{2,6}$. | 0 | |
| | | | | | 2 | $_bM_{1,2}$ |
| | | | | | 6 | $_bM_{5,6}$ |
| | | | 4 | $_bM_{5,6}$ | | |
| | 3 | $_cM_{3,4}$ | 0 | | | |
| | | | 3 | $_bM_{5,6}$ | | |
| | | | 4 | | | |

| Table 4.2 (Graph 4 (d)) | | | | | | | |
|---|---|---|---|---|---|---|---|
| outcome | operate | outcome | operate | outcome | operate | | |
| $_bM_{1,2}$ | 0 | $_cM_{3,4}$ | 0 | | | |
| | | | 3 | $_aM_{6,7}$ | | |
| | | | 4 | | | |
| | 1 | $_cM_{1,5}$ | 0 | | | |
| | | | 1 | $_aM_{6,7}$ | | |
| | | | 5 | | | |
| | 2 | $_cM_{2,3}$ | 0 | $_aM_{6,7}$ | | |
| | | | 2 | For 2_x_6, x can not be a,b,c, or it becomes case 3. And so $_dM_{2,6}$. | 0 | |
| | | | | | 2 | $_aM_{2,4}$ |
| | | | | | 6 | $_aM_{6,7}$ |
| | | | 3 | $_aM_{6,7}$ | | |

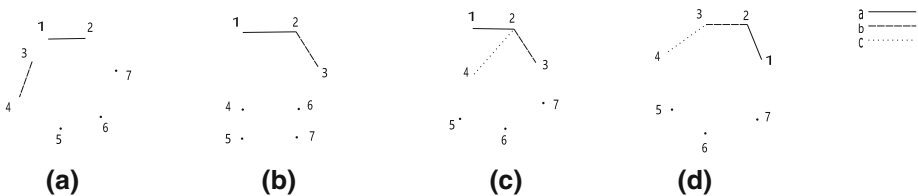

Graph 5

| Table 4.3 | | | | | |
|---|---|---|---|---|---|
| | outcome | operate | outcome | operate | outcome | operate |
| $_aM_{1,2}$ | 0 (Graph 5 (a)) | $_bM_{3,4}$ | 0 | | | |
| | | | 3 (If the outcome is 3 or 4, WLG, assume that it is 3.) | WLG, 3 x 5, x≠a. If x=b, then 5 is excluded; If x≠b, then $_cM_{3,5}$. | | |
| | 2 (If the outcome is 1 or 2, WLG, assume that it is 2.) | $_bM_{3,4}$ | 0 (Graph 5 (b)) | $_cM_{4,5}$ (If it is not case 4.1.) | | |
| | | | 2 (Graph 5 (c)) | $_cM_{2,4}$ | 0 | |
| | | | | | 2 | If there exists 2 x i, i≥5, x=a,b,c, then i is excluded; If not, then $_dM_{2,5}$. |
| | | | | | 4 | If there exists 4 c i, for i≥5, then i is excluded; If 4 x i, i≥5 implies that x=a,b, then it is case 4.1; If it is not the above cases, then $_dM_{4,5}$. |
| | | | 3 (Graph 5 (d)) | $_cM_{3,4}$ | 0 | |
| | | | | | 3 | If there exists 3 x i, for i≥5, x=b,c, then i is excluded; If 3 x i, i≥5 implies that x=a, then it is case 4.1; If it is not the above cases, then $_dM_{3,5}$. |
| | | | | | 4 | If there exists 4 c i, for i≥5, then i is excluded; If 4 x i, i≥5 implies that x≠a,b,c, then $_dM_{4,5}$; If it is not the above cases, then it is case 4.2. |